# Deterministic Formulization of SNR for Wireless Multiuser DS-CDMA Networks

Syed S. Rizvi and Khaled M. Elleithy
Computer Science and Engineering Department
University of Bridgeport
Bridgeport, CT 06601
{srizvi, elleithy}@bridgeport.edu

Aasia Riasat
Department of Computer Science
Institute of Business Management
Karachi, Pakistan 78100
aasia.riasat@iobm.edu.pk

*Abstract*—**Wireless Multiuser receivers suffer from their relatively higher computational complexity that prevents widespread use of this technique. In addition, one of the main characteristics of multi-channel communications that can severely degrade the performance is the inconsistent and low values of SNR that result in high BER and poor channel capacity. It has been shown that the computational complexity of a multiuser receiver can be reduced by using the transformation matrix (TM) algorithm [4]. In this paper, we provide quantification of SNR based on the computational complexity of TM algorithm. We show that the reduction of complexity results high and consistent values of SNR that can consequently be used to achieve a desirable BER performance. In addition, our simulation results suggest that the high and consistent values of SNR can be achieved for a desirable BER performance. The performance measure adopted in this paper is the consistent values of SNR.**

*Keywords*—*Computational complexity, DS-CDMA, wireless multiuser receivers, signal to noise ratio*

I. INTRODUCTION

From the design standpoint, for a given modulation and the coding scheme there is a one to one correspondence between the bit error rate (BER) and the signal-to-noise ratio (SNR). From the user standpoint, SNR is not the favorite criterion for the performance evaluation of digital communication links, because the user measures the quality of a system by the number of errors in the received bits and prefers to avoid the technical detail of modulation or coding. However, using received SNR rather than BER will allow us to relate our performance criteria to the required transmitted power, which is very important for battery-operated wireless operations. Using SNR rather than BER has two advantages. First, SNR is the criterion used for accessing both digital and analog modulation techniques. Second, SNR is directly related to the transmitted power, which is an important design parameter.

A significant amount of efforts have been made in order to achieve high values of SNR [3, 5]. However, none of these methods relate the complexity of multiuser receivers for achieving high SNR values. On the other hand, the TM algorithm is a low complexity, but synchronous transmission technique that is able to reduce the number of computations performs by a multiuser receiver for signal detection [4]. The TM algorithm therefore provides fast multiuser signal detection which can be further used to achieve high SNR values. The contribution of this research work is the quantification of SNR using the TM algorithm proposed by Rizvi et. [4]. At high SNR values, the error rate for multi channel can be reduced as well the capacity of the channel can be well approximated.

Multiuser receivers can be categorized in the following two forms: optimal maximum likelihood sequence estimation (MLSE) receivers and suboptimal linear and nonlinear receivers. Non-linear multiuser receiver involves the estimation and reconstruction of MAI [6] seen by each user with the objective of canceling it from the received signal. The two well known implementations of this mechanism are SIC and PIC. In interference cancellation, MAI is first estimated and then subtracted from the received signal [1, 7]. On the other hand, linear multiuser receivers apply a linear transformation to an observation vector, which serves as soft decision for the transmitted data. Recently, Ottosson and Agrell [2] proposed a new ML receiver that uses the neighbor descent (ND) algorithm. They implemented a linear iterative approach using the ND algorithm to locate the region where the actual observations belong. The linearity of their iterative approach increases noise components at the receiving end. Due to the enhancement in the noise components, the SNR and BER of ND algorithm is more affected by the MAI. Table 1, reported from [8], highlights the assumed knowledge for the computational complexity of a CDMA based multiuser receiver. Table I shows that different receivers distinguish themselves with respect to the requirement of the desired knowledge as well as the implementation complexity.

Verdu [1] proposed the optimum multiuser detector for asynchronous systems. The complexity of multiuser receiver grows exponentially in an order of $O(2)^K$, where $K$ is the number of active users. Recently, [2] proposed a ML receiver that uses the neighboring decent (ND) algorithm with an iterative approach to locate the regions. The linearity of the iterative approach increases noise components at the receiving end. The TM algorithm [4] observes the coordinates of the constellation diagram to determine the location of the transformation points. Since most of the decisions are correct, the TM algorithm can reduce the number of computations by using the transformation matrices only on those coordinates which are most likely to lead to an incorrect decision.

TABLE I. COMPLEXITY REQUIREMENTS OF DETECTION ALGORITHMS FOR DS-CDMA SYSTEMS

| Receivers | Signature of Desired User | Signature of Interference | Timing of Desired User | Timing of Interferers | Relative Amplitude | Training Sequence |
|---|---|---|---|---|---|---|
| **Conventional and Rake** | YES | NO | YES | NO | YES | NO |
| **Linear ZF** | YES | YES | YES | YES | NO | NO |
| **Linear MMSE** | YES | YES | YES | YES | YES | NO |
| **SIC and PIC** | YES | YES | YES | YES | YES | NO |
| **Trained Adaptive MMSE** | NO | NO | YES | NO | NO | YES |
| **Blind Adaptive MMSE** | YES | NO | YES | NO | NO | NO |

## II. PROPOSED TRANSFORMATION MATRIX ALGORITHM

We consider a synchronous DS-CDMA system as a linear time invariant (LTI) channel. In a LTI channel, the probability of variations in the interference parameters, such as the timing of all users, amplitude variation, phase shift, and frequency shift, is extremely low. This property makes it possible to reduce the overall computational complexity at the receiving end. Our TM technique utilizes the complex properties of the existing inverse matrix algorithms to construct the transformation matrices and to determine the location of the TPs that may occur in any coordinate of the constellation diagram. The individual TPs can be used to determine the average computational complexity.

The system may consist of $K$ users. User $k$ can transmit a signal at any given time with the power of $W_k$. With the binary phase shift keying (BPSK) modulation technique, the transmitted bits belong to either +1 or -1, (i.e., $b_k \in \{\pm 1\}$). The cross correlation can be reduced by neglecting the variable delay spreads, since these delays are relatively small as compared to the symbol transmission time. In order to detect signals from any user, the demodulated output of the low pass filter is multiplied by a unique signature waveform assigned by a pseudo random number generator. It should be noted that we extract the signal using the match filter followed by a Viterbi algorithm. The optimum multiuser receiver exists and permits to relax the constraints of choosing the spreading sequences with good correlation properties at a cost of increased receiver complexity.

### A. Description of Transformation Matrix Algorithm

According to original Verdu's algorithm, the outputs of the matched filter $y_1(m)$ and $y_2(m)$ can be considered as a single output $y(m)$. In order to minimize the noise components and to maximize the received demodulated bits, we can transform the output of the matched filter, and this transformation can be expressed as follows: $y(m) = Tb + \eta$ where $T$ represents the TM, $b_k \in \{\pm 1\}$ and $\eta$ represents the noise components. In addition, if the vectors are regarded as points in K-dimensional space, then the vectors constitute a constellation diagram that has $K$ total points.

The constellation diagram can be mathematically expressed as: $X = \{Tb\}$ where $b \in \{-1, +1\}$ and $X$ represents the collective computational complexity of a multiuser receiver.

The preceding equation is fundamental to the proposed algorithm. According to the detection rule, the constellation diagram can be partitioned into $2^K$ lines (where the total possible lines in the constellation diagram can be represented as $f$) that can only intersect each other at the following points: $X = \{Tb\}_{b \in \{-1, 1\}^K} \backslash f$.

Fig. 1 shows the constellation diagram that consists of three different vectors (lines) with the original vector '$X$' that represents the collective complexity of the receiver. $Q$, $R$, and $S$ represent vectors or TP within the coverage area of a cellular network (see Fig. 1). In addition, $Q^¬$, $R^¬$, and $S^¬$ represent the computational complexity of each individual TP. In order to compute the collective computational complexity of the optimum receiver, it is essential to determine the complexity of each individual TP.

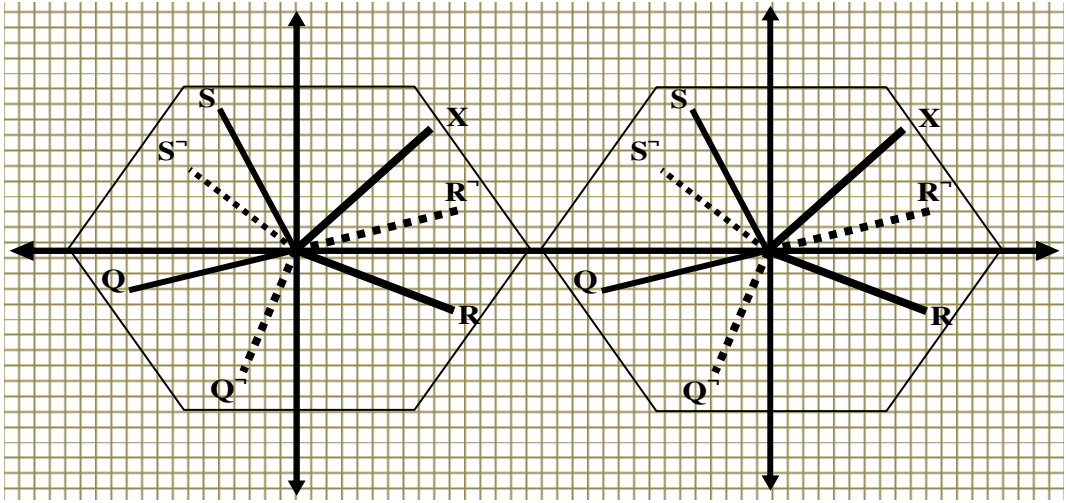

Figure 1. A constellation diagram consisting of three different vectors

The computational complexity of each individual TP is represented by $X^\neg$ of the TP which is equal to the collective complexity of $Q^\neg$, $R^\neg$, and $S^\neg$. In order to derive the value of the original vector $X$, we need to perform the following derivations. We consider the original vector with respect to each transmitted symbol or bit.

$$X^\neg Q = Xi^\neg = \langle XQ_i + XR_j + XS_K \rangle i^\neg =$$
$$\langle XQ_i i^\neg + XR_j i^\neg + XS_K i^\neg \rangle$$
$$X^\neg R = Xj^\neg = \langle XQ_i + XR_j + XS_K \rangle j^\neg =$$
$$\langle XQ_i j^\neg + XR_j j^\neg + XS_K j^\neg \rangle$$
$$X^\neg S = Xk^\neg = \langle XQ_i + XR_j + XS_K \rangle k^\neg =$$
$$\langle XQ_i k^\neg + XR_j k^\neg + XS_K k^\neg \rangle$$

The following equation can be derived from the above system:

$$\left\langle \begin{matrix} X^\neg Q \\ X^\neg R \\ X^\neg S \end{matrix} \middle| \begin{matrix} i^\neg \\ j^\neg \\ k^\neg \end{matrix} \right\rangle = \begin{bmatrix} i(i^\neg) & j(i^\neg) & k(i^\neg) \\ i(j^\neg) & j(j^\neg) & k(j^\neg) \\ i(k^\neg) & j(k^\neg) & k(k^\neg) \end{bmatrix} \begin{bmatrix} XQ \\ XR \\ XS \end{bmatrix} \quad (1)$$

Equation (1) represents the following: $QRS$ with the unit vectors $i$, $j$, and $k$, and $X^\neg Q$, $X^\neg R$, and $X^\neg S$ with the inverse of the unit vectors $i^\neg$, $j^\neg$, and $k^\neg$. The second matrix on the right hand side of (1) represents **b**, where as the first matrix on the right hand side of (1) represents the actual TM. The TM from the global reference points to a particular local reference point can now be derived from (1):

$$\left\langle \begin{matrix} X^\neg Q \\ X^\neg R \\ X^\neg S \end{matrix} \middle| \begin{matrix} i^\neg \\ j^\neg \\ k^\neg \end{matrix} \right\rangle = T_{L/G} \begin{bmatrix} XQ \\ XR \\ XS \end{bmatrix} \quad (2)$$

Equation (2) can also be written as:

$$T_{L/G} = \begin{bmatrix} ii^\neg & ji^\neg & ki^\neg \\ ij^\neg & jj^\neg & kj^\neg \\ ik^\neg & jk^\neg & kk^\neg \end{bmatrix} \quad (3)$$

In (3), the dot products of the unit vectors of the two reference points are in fact the same as the unit vector of the inverse TM of (2). We need to compute the locations of the actual TP described in (2) and (3). Let the unit vectors for the local reference point be:

$$i^\neg = [T_{11}i, T_{12}j, T_{13}k]$$
$$j^\neg = [T_{21}i, T_{22}j, T_{23}k] \quad (4)$$
$$k^\neg = [T_{31}i, T_{32}j, T_{33}k]$$

Since, $i^\neg (i + j + k) = i^\neg$, where $(i + j + k) = 1$. The same argument is true for the rest of the unit vectors. Therefore, (4) can be rewritten as:

$$i^\neg = [T_{11}, T_{12}, T_{13}]$$
$$j^\neg = [T_{21}, T_{22}, T_{23}] \quad (5)$$
$$k^\neg = [T_{31}, T_{32}, T_{33}]$$

By substituting the values of $\vec{i}$, $\vec{j}$, and $\vec{k}$ from (5) into (3), we obtain

$$T_{L/G} = \begin{bmatrix} i\langle T_{11}i, T_{12}j, T_{13}k \rangle & j\langle T_{11}i, T_{12}j, T_{13}k \rangle & k\langle T_{11}i, T_{12}j, T_{13}k \rangle \\ i\langle T_{21}i, T_{22}j, T_{23}k \rangle & j\langle T_{21}i, T_{22}j, T_{23}k \rangle & k\langle T_{21}i, T_{22}j, T_{23}k \rangle \\ i\langle T_{31}i, T_{32}j, T_{33}k \rangle & j\langle T_{31}i, T_{32}j, T_{33}k \rangle & k\langle T_{31}i, T_{32}j, T_{33}k \rangle \end{bmatrix}$$

$$T_{L/G} = \begin{bmatrix} T_{11}, T_{12}, T_{13} \\ T_{21}, T_{22}, T_{23} \\ T_{31}, T_{32}, T_{33} \end{bmatrix} \quad (6)$$

Substituting $T_{L/G}$ from (6) into (2), yields

$$\begin{pmatrix} \langle X\vec{}Q \rangle \\ \langle X\vec{}R \rangle \\ \langle X\vec{}S \rangle \end{pmatrix} = \begin{bmatrix} T_{11} & T_{12} & T_{13} \\ T_{21} & T_{22} & T_{23} \\ T_{31} & T_{32} & T_{33} \end{bmatrix} \begin{bmatrix} XQ \\ XR \\ XS \end{bmatrix} \quad (7)$$

Equation (7) corresponds to the following standard equation that used for computing the computational complexity at the receiving end: $X = Tb$ where $b \in \{-1, +1\}^k$.

If the target of one transformation $(U:Q \to R)$ is the same as the source of other transformation $(T:R \to S)$, then we can combine two or more transformations and form the following composition: $TU: Q \to S$, $TU(Q) = T[U(Q)]$. This composition can be used to derive the collective computational complexity at the receiving end using (7). Since we assumed that the transmitted signals are modulated using BPSK which can at most use 1 bit out of two bits (i.e., $b_k \in \{\pm 1\}$), consider the following set of TP to approximate the number of demodulated received bits that need to search out by decision algorithm:

$$\begin{bmatrix} y(m) \\ y(m+1) \\ \vdots \\ \\ \\ y(K) \end{bmatrix} = \begin{bmatrix} Tb(0) & Tb(-1) & 0 & \cdots & 0 \\ Tb(1) & Tb(0) & Tb(-1) & \cdots & 0 \\ 0 & Tb(1) & Tb(0) & \cdots & 0 \\ & \ddots & \ddots & & \\ & & & \ddots & \\ \vdots & \vdots & \ddots & & \vdots \\ 0 & \cdots & \cdots & Tb(1) & Tb(0) \end{bmatrix} + \begin{bmatrix} \eta(m) \\ \eta(m+1) \\ \\ \vdots \\ \\ \eta(m+k) \end{bmatrix} \quad (8)$$

Equation (8) is derived using our fundamental equation of TM (i.e., $y = Tb + \eta$). Our approach is to assume terms $\eta(m)$ and $\eta(m+k)$ in (7) not equal to zero. This condition is fulfilled by periodically inserting a nonzero-energy bit in the information bit sequence. Therefore, the interference due to the cross-correlation of the actual symbols with the past and future symbols in the asynchronous channels can be accounted.

Using (7), a simple matrix addition of the received demodulated bits can be used to approximate the number of most correlated TP. The entire procedure for computing the number of demodulated bits that need to be searched out by the decision algorithm can be used to approximate the number of most correlated signals for any given set of TP. This is because we need to check whether or not the TP are closest to either (+1, +1) or (-1, -1). The decision regions or the coordinates where the TP lie for (+1, +1) and (-1, -1) are simply the corresponding transformation matrices that store the patterns of their occurrences. If the TP do not exist in the region of either (+1, +1) or (-1, -1), then it is just a matter of checking whether the TP are closest to (+1, -1) or to (-1, +1).

The minimum search performed by the decision algorithm is conducted if the TP exist within the incorrect region. Since the minimum search saves computation by one degree, the decision algorithm has to search at least $4^k$ demodulated bits. This implies that the total number of demodulated bits that need to be searched out by the decision algorithm can not exceed by $5^K - 4^K$. Thus, the total number of most correlated pairs has an upper bound of $5^K - 4^K$.

Since most of the decisions are correct, we can reduce the number of computations by using the transformation matrices only on those coordinates that are most likely to lead to an incorrect decision. Thus, this greatly reduces the unnecessary processing required to make a decision about the correct region. Thus, the number of received demodulated bits that need to be searched out can be approximated as $5^K - 4^K$.

The computational complexity of any multiuser receiver can be quantified by its time complexity per bit [6]. The collective computational complexity of the proposed algorithm is achieved after performing the TM sum. This implies that both quantities $T$ and $b$ from our fundamental equation can be computed together and the generation of all the values of the demodulated received bits $b$ can be done through the sum of the actual TM $T$ that approximately takes $O(5/4)^k$ operations with an asymptotic constant. Using the Newton approximation method given in MATLAB, we can directly come to an approximation of $O(5/4)^k$. The computational complexity of the proposed algorithm is not polynomial in the number of users, instead the number of operations required to maximize the demodulation of the transmitted bits and to choose an optimal value of b is $O(5/4)^k$, and therefore the time complexity per bit is $O(5/4)^k$.

(IJCSIS) International Journal of Computer Science and Information Security
Vol. 3, No. 1, 2009III. PROPOSED QUANTIFICATION OF SIGNAL-TO-NOISE RATION (SNR)

In this section, we derive an expression to provide quantification of SNR for the signals received at the DS-CDMA multiuser receiver. The reduced complexity of the TM algorithm provides faster detection rate. The faster detection rate results high and consistent values of SNR. Once we determine the values of SNR, we can relate them to the BER performance and the channel capacity approximation for a wireless multiuser receiver.

MAI causes the SNR degradation resulting in a degraded SNR performance for a particular value of $E_b/N_o$. We present that due to the reduced complexity, the SNR performance of the TM algorithm would remain consistent in terms of the desired values even for a large value of $K$. This consistency in SNR performance yields an optimal BER performance.

*A. System Model and Key Assumptions*

Our fundamental assumption is that the system is linear time invariant (LTI) which leads us to the fact that the transmitted signals experience no deep fades. Due to the linearity and time invariant properties of the system, we can ignore the phase shift, and deep fades. In other words, the overall SNR of the received signals has a slow convergence rate compared to the convergence rate of the BER.

*B. Proposed Formulization for SNR*

Consider the following assumptions for an AWGN channel:

(a) $\aleph$ *represents the computational complexity that belongs to a certain coverage area.*

(b) *SNR (we represent SNR by $\gamma$) is uniformly distributed among all the active user's signals with respect to computational complexity.*

(c) *A certain cellular coverage area has K users.*

Based on these above assumptions, we can give the following hypothesis:

$$\aleph_i \in \{\aleph_1, \aleph_2, \aleph_3, \ldots\ldots\ldots\ldots, \aleph_K\} \tag{9}$$

where $\aleph_1, \aleph_2, \aleph_3, \ldots\ldots\ldots\ldots\aleph_K$ indicates the indicates the computational complexity-domain and

$$h_i \in \{h_1, h_2, h_3, \ldots\ldots\ldots\ldots, h_K\} \tag{10}$$

where $h_1, h_2, h_3, \ldots\ldots\ldots\ldots, h_K$ indicates the user-domain.

Complexity-domain can be considered as a simple data structure for storing the patterns of occurrences of all active users. User-Domain is the number of active users present in the certain coverage area of a cellular network. The collective computational complexity can be expressed as:

$$\aleph = \sum_{i=1}^{K} \aleph_i \ where \ i = 1, 2, \ldots\ldots, K \tag{11}$$

Since each user has $h_{th}$ part of the computational complexity such as: $h_1 \in \aleph_1, h_2 \in \aleph_2, \ldots\ldots, h_K \in \aleph_K$.

This implies that each active user in a certain area of a cellular network has an average of $\aleph/K$ computational complexity. Since SNR is uniformly distributed among all the user's signals at the receiving end, each user experiences an average of $\gamma/K$ SNR. Therefore, this argument leads us to:

$$K/\aleph = C^{-1} - C^{-1}(\gamma/\aleph) = \left[1 - (\gamma/\aleph)C^{-1}\right] \tag{12}$$

where $C$ in (12) represents the normalization factor, $K/\aleph$ is the inverse of the computational complexity, and $\gamma/\aleph$ represents the SNR with respect to average computational complexity.

Equation (12) can be interpreted that the inverse of computational complexity equals to the difference between the inverse-normalization factor and the product of the inverse-normalization factor and SNR with respect to the collective computational complexity. The main objective of (12) is to make sure that we should get maximum positive values of SNR for most of the values of K.

*C. Proof for $\gamma/\aleph$*

If the previous assumptions are valid for an AWGN channel, the following approximation must be true for both the complexity and the user domains:

$$\aleph/K \xrightarrow{approximation} C + \gamma/K \tag{13}$$

We present our hypothesis that the difference between the average computational complexity and the average SNR should equal to the normalization factor. The main objective of (13) is to get maximum positive values of SNR for most of the values of *K*. Equation (5) can also be written as:

$$(\aleph/K) - (\gamma/K) = C \tag{14}$$

Based on (14), we can write the following equation:

2929

$$\gamma/\aleph = 1 - C(K/\aleph) \quad (15)$$

Since the right hand side of (15) represents the inverse of the average computational complexity with the normalization factor, the number of required operations can not be less than zero. It should be noted that the right hand side of (15) always gives us a positive value of SNR for any value of $K$ which is greater than 10. Equation (15) can also be rewritten as:

$$K/\aleph = C^{(-1)}[1 - \gamma/\aleph] \quad (16)$$

Using the complexity and the user domain, we can make an argument that the inverse of an average SNR should be at least greater than zero. This argument guarantees that the system does not work with a non positive value of SNR. In other words, the inverse of the average SNR should equal to the difference of the normalization factor and the inverse of the average computational complexity. Recall (12):

$$K/\aleph = C^{-1}[(\aleph - \gamma)/\aleph] \triangleq \gamma = (\aleph - CK) \quad (17)$$

Equation (17) represents SNR by determining the difference between the power of the transmitted signal from the computational complexity-domain and the number of users from the user-domain. Equation (17) can also be used to compute the values of SNR in an ideal situation only if MAI does not affect the received signals by *K-1* users. However, in a practical DS-CDMA system, this assumption does not exist. Therefore, we should consider that the variations in the network load for an AWGN channel introduces the presence of variance (we represent variance by $\Phi^2$) that represents MAI.

The selection of variance is entirely dependent on the network load. The variance is a linear function of the active users ($K$) and it should increase as we increase the value of $K$. In order to compute the values of SNR, we need to change the linear quantity into decibels (dB) by multiplying it to the base-10 logarithmic function as well as with the variance. This leads us to the following expression for SNR:

$$\gamma = 10\Phi^2 \log_{10}(\aleph - CK) \quad (18)$$

We use the values of variance in our simulation that represents MAI with respect to *K*.

IV. EXPERIMENTAL VERIFICATION AND SIMULATION RESULTS

Fig. 2 shows the logical diagram of a cellular system that uses synchronous DS-CDMA system. We assume that all the users among the cellular area communicate using AWGN channels through one or more base stations. Because of AWGN multiple channels, the symbol duration of the transmitted signal is much larger than the delay spread which avoid inter-symbol interference. Therefore the uplink (from user to BS) model is based on synchronous DS CDMA system with multiple path channels and the presence of AWGN with zero mean and a varying amount of variance.

The choice of variance depends on the number of active users present in the coverage area of a cellular network. Furthermore, we assume that the transmission power of each user is tightly controlled (which is a usual thing for wireless applications) by the central entity of a coverage area such as a central base station (BS) or an access point (AP). This implies that the central entity (BS/AP) of a coverage area receives uniform-power-signals and they remain same throughout the total communication time.

It has been shown that the SNR degradation depends on the number of users, *K*, [4]. An increase in *K* would degrade the performance because it would increase the cross correlation between the received signals from all the users (i.e., *K-1* users). Mathematically, we can express this as: $K \propto MAI \propto high\ BER \propto 1/SNR$. This shows that a slight increase in *K* would degrade the SNR performance that consequently increases the BER. However, a large increase in value of *K* forces MAI to reach its peak value that limits the divergence of SNR for the TM algorithm.

Three different types of detection algorithms are investigated, which are the original ML algorithm, reduced ND algorithm, and the proposed TM algorithm [4]. The following is the description of the parameters that we use for two different scenarios: (i) *lightly-loaded network* where *K* starts from 2 to 50 and (ii) *heavily-loaded network* where K starts from 2 to 100. LTI synchronous DS-CDMA over an AWGN channel with small variation in $\Phi^2$ are used.

In order to compare the SNR performance of the proposed algorithm with the other multiuser detection algorithms, we use a same constant value with their asymptotic computational complexities that does not make an exception for any one of the investigated algorithms. In our simulation for both scenarios, we use one (i.e., $C = 1$) as a normalization factor that remains same for all the investigated algorithms.

The choice of a small value of $\Phi^2$ is entirely based on the load of the coverage area (*K*) and it is selected through a random process for a certain range of users. For a lightly loaded network, we expect that the value of variance ($\Phi^2$) may vary from 0.6 to 0.9 and for a heavily-loaded network; the value of variance may vary from 0.1 to 1. Since the proposed algorithm detects transmitted signals by using complex properties of inverse matrix algorithm that observes the coordinates of the constellation diagram to determine the location of the corresponding transformation points, it is more likely that the value of variance is extremely small for both lightly-loaded as well as heavily-loaded networks. Furthermore, all signals transmit at the same bit rate and all signals receive with the same power (i.e., perfect power control).

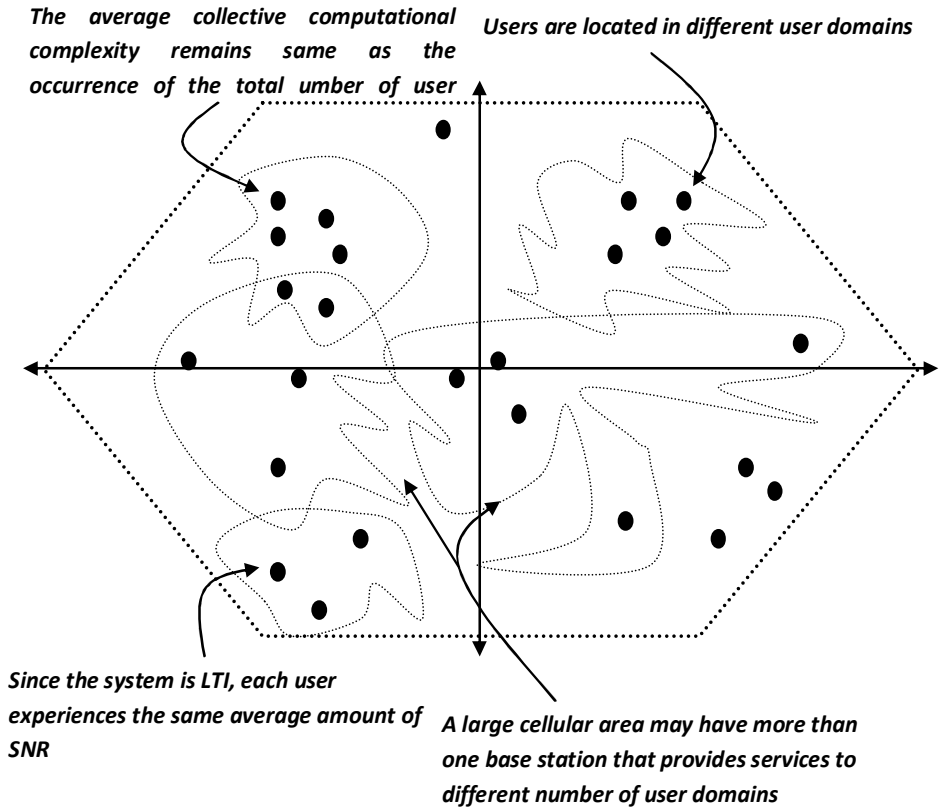

Figure 2. Logical Diagram of a Cellular network

*For* lightly-loaded network, (2<K<50) whereas for heavily-loaded network (2<K<100). LTI synchronous DS-CDMA over an AWGN channel with small variation in $\Phi^2$ are used. The choice of a small value of variance is entirely based on the value of *K* and it is selected through a random process.

### A. *Performance Evaluation for Lightly Loaded Networks*

Fig. 3 shows one of the possible cases of a lightly-loaded network where 22 active users transmit BPSK modulated signals. For a small value of *K*, the proposed TM algorithm achieves approximately 6.5 dB of SNR where as the ND and the ML algorithms give 5.8 and 5.5 dB, respectively.

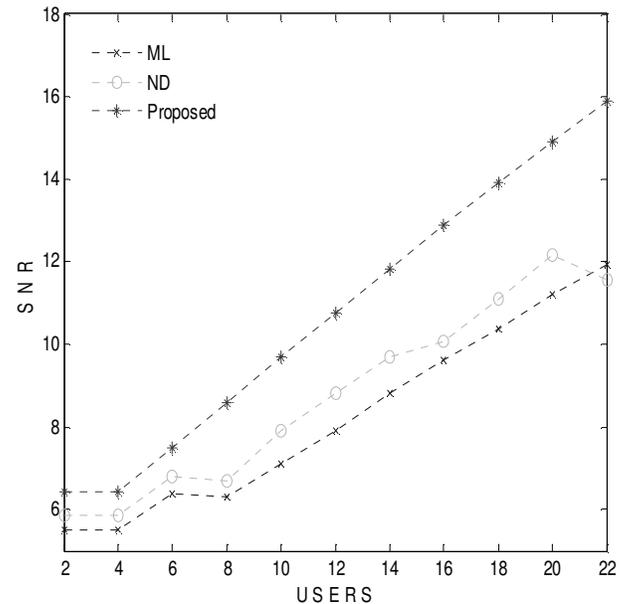

Figure 3. Approximate values of SNR (dB) versus number of users (K=22) with a random amount of variance for a synchronous system in an AWGN channel.

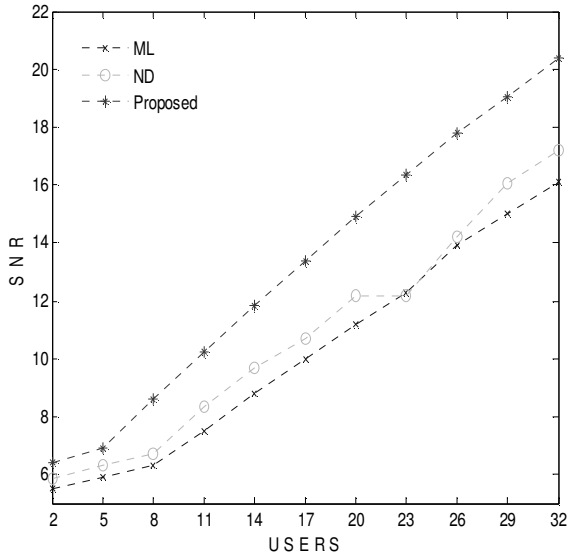

Figure 4. Approximate value of SNR (dB) versus number of users (K =32) with a random amount of variance for a synchronous system in an AWGN channel.

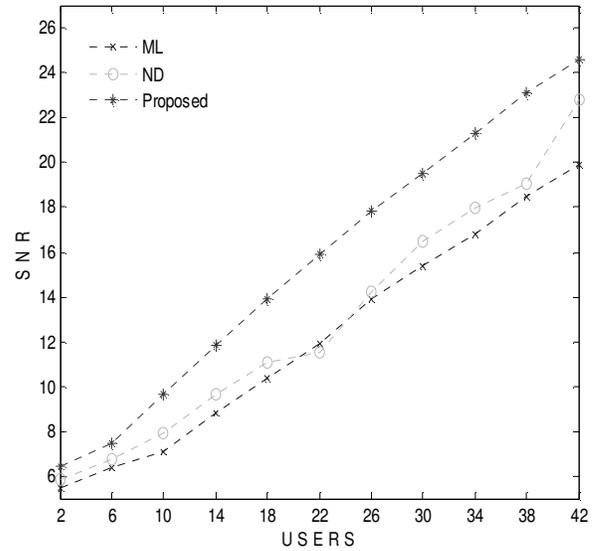

Figure 5. Approximate value of SNR (dB) versus number of users (K =42) with a random amount of variance for a synchronous DS-CDMA system in a Gaussian channel.

This implies that a slight increase in the value of $K$ forces the TM algorithm to give an acceptable value of SNR that can be used to achieve a satisfactory BER performance at least for a voice communication network. This can be seen in Fig. 4 that the TM algorithm has more rapid divergence with respect to the number of users than the ND and the ML algorithms. The divergence in SNR is directly proportional to the convergence in BER performance. In addition, it can be clearly observed in Fig. 4 that the linear increase in SNR for the TM algorithm is more uniform and smoother over the ND and the ML algorithms.

Furthermore, the importance of variance can not be ignored, since Figures 3 and 4 clearly depict that a random amount of variance is more affected on the ND and the ML algorithms than on the proposed algorithm. This is because both ML and ND algorithms have comparatively larger complexity-domains which take more time to perform required iterations to detect the received signals and thus give more time to variance to effect comprehensively on the received SNR. The degradation in SNR due to variance can be seen in Figures 5 and 6 when $K = 42$ and $K = 52$, respectively. Moreover, for a lightly-loaded network, it can be expected that the selection of variance within the specified range does not meet the threshold value. In other words, the random amount of variance is more likely unstable for a lightly-loaded network than in a heavily-loaded network and thus may cause a serious degradation in the values of SNR.

### B. Performance Evaluation for Heavily Loaded Networks

For heavily-loaded case, we consider a cellular network that consists of approximately 2 to 100 active users. Fig. 7 and 8 is one of the examples of a heavily-loaded case where 72 to 102 active users transmit signals through the central entity of the network.

Fig. 7 shows that the linear increase in SNR is consistent not only for a lightly-loaded network but also for a heavily-loaded network. However, this can also be noticed from Figures 7 and 8 that as the number of users increase in the system, the differences between the SNR values for the proposed algorithm and the other two ML and the ND algorithms become wider. From Fig. 7, the proposed algorithm gives approximately 36 dB for K = 72 which is more than what we expect to achieve for an optimal BER performance. In addition to that, the random amount of variance is more affected on the SNR values in a heavily-loaded case than in a lightly-loaded case.

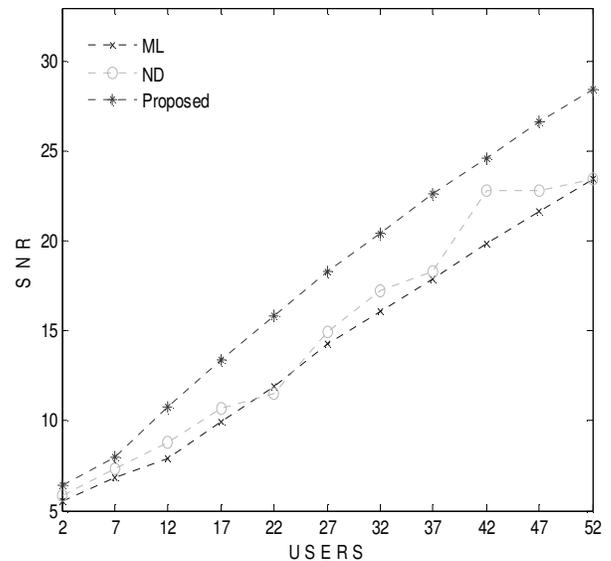

Figure 6. Approximate value of SNR (dB) versus number of users (K =52) with a random amount of variance for a synchronous DS-CDMA system in a Gaussian channel.

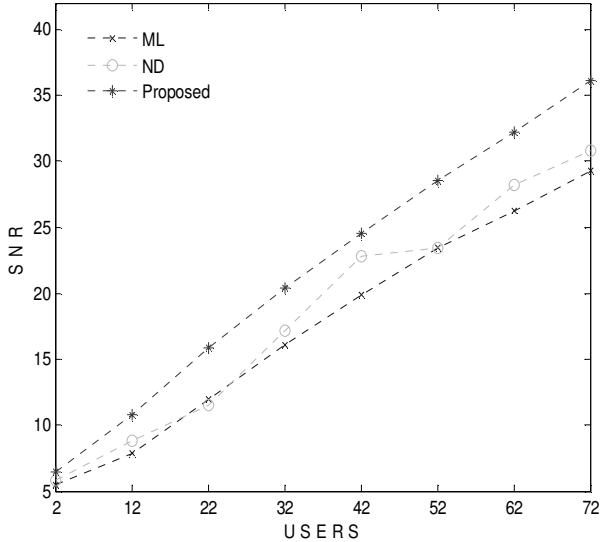

Figure 7. Approximate value of SNR (dB) versus number of users (K =72) with a random amount of variance for a synchronous DS-CDMA system in a Gaussian channel.

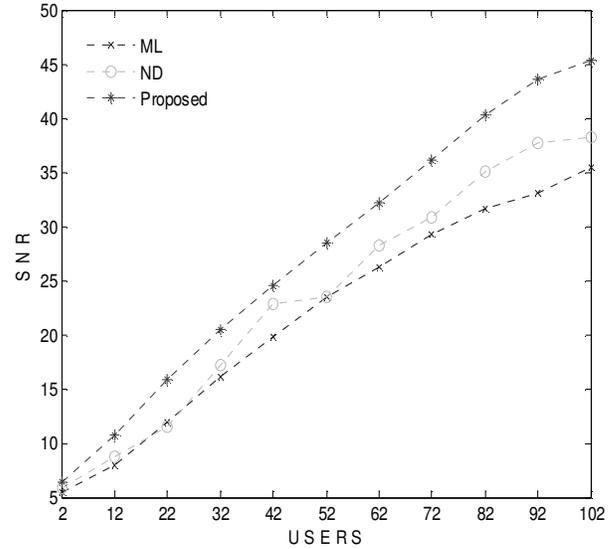

Figure 8. Approximate value of SNR (dB) versus number of users (K =102, heavily-loaded network) with a random amount of variance for a synchronous system in an AWGN channel.

In Fig. 8, it can be seen that the ND algorithm comparatively gets high values of SNR than the ML algorithm in a heavily-loaded network (typically when $K > 55$) when compare to a lightly-loaded network. This is because the computational complexity for a heavily-loaded case is much greater than the computational complexity for a lightly-loaded case that forces both ML and ND algorithms to minimize the factor of divergence and hence maximize the factor of convergence. Since we assume that the selection of variance is random within the specified range, it remains stable after a certain value of K that limits the divergence of SNR.

Another important point that can be observed from Fig. 8 is that the graph for the proposed algorithm converges to approximately 45 dB after 100 users and only a slight increase in the value of SNR can be expected for very large values of $K$. This is also essential for achieving an acceptable performance, since crossing the threshold value of SNR might degrade the overall system performance. In other words, after a certain value of K, the MAI reaches to its peak value that limits the divergence of the SNR curve for the proposed algorithm.

## V. CONCLUSION

In this paper, we presented the quantification of SNR based on the TM algorithm. We have shown that the reduction in the computational complexity of a multiuser receiver can be used to achieve high and consistent values of SNR. The simulation results suggest that due to a low complexity domain, the SNR performance of the TM algorithm is more uniform and smoother over the other well known algorithms. For the future work, it will be interesting to implement the proposed approach for asynchronous systems to achieve desirable BER performance and approximate the capacity of a multi channel.

array systems," Wireless Networks , Kluwer Academic Publishers. Manufactured in The Netherlands, Vol. 9, Issue 4, pp. 373-378, July 2003.

[8] C. Piero, *Multiuser Detection in CDMA Mobile Terminals*. Artech House, Inc., 2002.

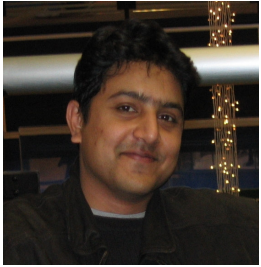

**Syed S. Rizvi** is a Ph.D. student of Computer Science and Engineering at University of Bridgeport. He received a B.S. in Computer Engineering from Sir Syed University of Engineering and Technology and an M.S. in Computer Engineering from Old Dominion University in 2001 and 2005, respectively. In the past, he has done research on bioinformatics projects where he investigated the use of Linux based cluster search engines for finding the desired proteins in input and outputs sequences from multiple databases. For last three year, his research focused primarily on the modeling and simulation of wide range parallel/distributed systems and the web based training applications. Syed Rizvi is the author of 68 scholarly publications in various areas. His current research focuses on the design, implementation and comparisons of algorithms in the areas of multiuser communications, multipath signals detection, multi-access interference estimation, computational complexity and combinatorial optimization of multiuser receivers, peer-to-peer networking, network security, and reconfigurable coprocessor and FPGA based architectures.

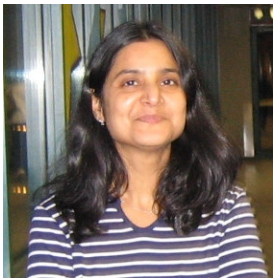

**AASIA RIASAT** is an Associate Professor of Computer Science at Collage of Business Management (CBM) since May 2006. She received an M.S.C. in Computer Science from the University of Sindh, and an M.S in Computer Science from Old Dominion University in 2005. For last one year, she is working as one of the active members of the wireless and mobile communications (WMC) lab research group of University of Bridgeport, Bridgeport CT. In WMC research group, she is mainly responsible for simulation design for all the research work. Aasia Riasat is the author or co-author of more than 40 scholarly publications in various areas. Her research interests include modeling and simulation, web-based visualization, virtual reality, data compression, and algorithms optimization.

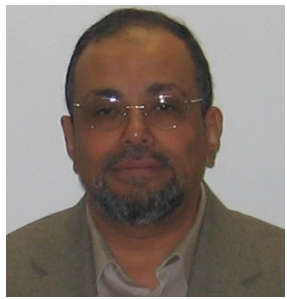

**KHALED ELLEITHY** received the B.Sc. degree in computer science and automatic control from Alexandria University in 1983, the MS Degree in computer networks from the same university in 1986, and the MS and Ph.D. degrees in computer science from The Center for Advanced Computer Studies at the University of Louisiana at Lafayette in 1988 and 1990, respectively. From 1983 to 1986, he was with the Computer Science Department, Alexandria University, Egypt, as a lecturer. From September 1990 to May 1995 he worked as an assistant professor at the Department of Computer Engineering, King Fahd University of Petroleum and Minerals, Dhahran, Saudi Arabia. From May 1995 to December 2000, he has worked as an Associate Professor in the same department. In January 2000, Dr. Elleithy has joined the Department of Computer Science and Engineering in University